\documentclass{iaus}

\usepackage{graphics,epsfig}

\title[Biological damage in stellar environments]
{Biological damage due to photospheric, \\
 chromospheric and flare radiation \\
 in the environments of main-sequence stars}

\author[M. Cuntz et al.]   
{Manfred Cuntz$^1$, Edward F. Guinan$^2$ \and Robert L. Kurucz$^3$}

\affiliation{
$^1$Department of Physics, University of Texas at Arlington, \\
Arlington, TX 76019-0059, USA \\ email: {\tt cuntz@uta.edu} \\ [\affilskip]
$^2$Department of Astronomy and Astrophysics, Villanova University, \\
Villanova, PA 19085, USA \\email: {\tt edward.guinan@villanova.edu} \\ [\affilskip]
$^3$Harvard-Smithsonian Center for Astrophysics, \\
Cambridge, MA 02138, USA \\email: {\tt rkurucz@cfa.harvard.edu}
}

\pubyear{2010}
\volume{264}  
\pagerange{xxx--xxx}
\jname{Planetary Systems as Potential Sites for Life}
\editors{A.C. Editor, B.D. Editor \& C.E. Editor, eds.}
\begin{document}

\maketitle

\begin{abstract}
  We explore the biological damage initiated in the environments of F, G, K,
  and M-type main-sequence stars due to photospheric, chromospheric and flare
  radiation.  The amount of chromospheric radiation is, in a statistical sense,
  directly coupled to the stellar age as well as the presence of significant
  stellar magnetic fields and dynamo activity.  With respect to photospheric
  radiation, we also consider detailed synthetic models, taking into account
  millions or hundred of millions of lines for atoms and molecules.
  Chromospheric UV radiation is increased in young stars in regard to all
  stellar spectral types.  Flare activity is most pronounced in K and M-type
  stars, which also has the potential of stripping the planetary atmospheres
  of close-in planets, including planets located in the stellar habitable zone.
  For our studies, we take DNA as a proxy for carbon-based macromolecules,
  guided by the paradigm that carbon might constitute the biochemical
  centerpiece of extraterrestrial life forms.  Planetary atmospheric
  attenuation is considered in an approximate manner.
\end{abstract}

\firstsection 

\section{Theoretical Approach}

The centerpiece of all life on Earth is carbon-based biochemistry. It has
repeatedly been surmised that biochemistry based on carbon may also play a
pivotal role in extraterrestrial life forms, if existent.
This is due to the pronounced advantages of carbon, especially compared to its
closest competitor (i.e., silicon), which include: its relatively high
abundance, its bonding properties, and its ability to form very large molecules
as it can combine with hydrogen and other molecules as, e.g., nitrogen and oxygen
in a very large number of ways (\cite[Goldsmith \& Owen 2002]{gol02}).

In the following, we explore the relative damage to carbon-based macromolecules
in the environments of late-type main-sequence stars using DNA as a proxy.
We focus on the effects of both photospheric and chromospheric radiation and
comment on the significance of flare activity.  Previous studies indicating
the importance of UV radiation concerning the suitability of extrasolar planets
for the existence of life and biological evolution has been given by, e.g.,
\cite{coc99}, \cite{gui02}, \cite{gui03}, \cite{rib05}, \cite{buc06}, and
\cite{gui09}.  With respect to the Sun-Earth system, a detailed investigation
about the UV effects on biological systems has been pursued by \cite{dif91}.
This line of studies also includes the detailed modeling of the atmospheric
attenuation of UV (especially due to an Earth-type ozone layer, e.g.,
\cite[Segura et al. 2003]{seg03}), photobiological effects and the
effects of UV on biomolecules, particularly DNA, and microorganisms.

\begin{figure*}
\vspace*{+3.5 cm}
\centering
\epsfig{file=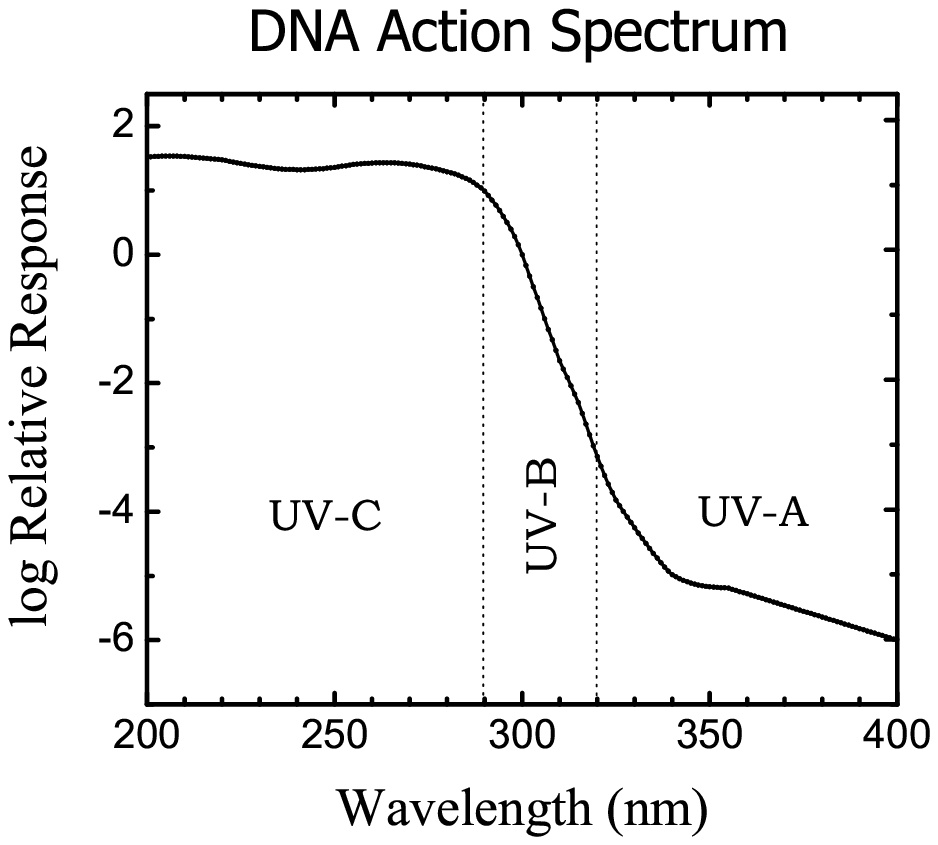,width=0.5\linewidth,height=0.45\linewidth}
\vspace*{-3.0 cm}
\caption{DNA action spectrum following \cite{hor95} and \cite{coc99}.
}
\end{figure*}

\begin{table}
\caption{Stellar Data and Habitable Zones}
\centering
\begin{tabular}{l c c c c c c c c}
\noalign{\smallskip}
\hline
\noalign{\smallskip}
Sp. Type &  $T_{\rm eff}$ &  $R_\ast$ & $L_\ast$ &
HZ-i (gen.) & HZ-i (con.) & HZ (E.) & HZ-o (con.) & HZ-i (gen.) \\
\noalign{\smallskip}
...   &  (K) & ($R_\odot$) & ($L_\odot$) & (AU) & (AU) & (AU) & (AU) & (AU) \\
\hline
\noalign{\smallskip}
F0~V  &   7200  &   1.62  &  6.33  &  1.826 &  2.251 &  2.516 &  3.222 &  3.710  \\
F5~V  &   6440  &   1.40  &  3.03  &  1.370 &  1.614 &  1.740 &  2.316 &  2.741  \\
G0~V  &   6030  &   1.12  &  1.49  &  1.002 &  1.152 &  1.220 &  1.657 &  1.991  \\
G5~V  &   5770  &   0.95  &  0.90  &  0.799 &  0.904 &  0.948 &  1.302 &  1.580  \\
K0~V  &   5250  &   0.83  &  0.47  &  0.604 &  0.665 &  0.685 &  0.961 &  1.188  \\
K5~V  &   4350  &   0.64  &  0.13  &  0.342 &  0.359 &  0.363 &  0.525 &  0.670  \\
M0~V  &   3850  &   0.48  &  0.05  &  0.207 &  0.213 &  0.213 &  0.313 &  0.407  \\
\hline
\noalign{\smallskip}
\multicolumn{9}{l}{Abbreviations: con. = conservative, gen. = general,
E. = Earth-equivalent} \\
\end{tabular}
\end{table}

We consider the radiative effects on DNA by applying a DNA action spectrum
(\cite[Horneck 1995]{hor95}) that shows that the damage is strongly
wavelength-dependent, increasing by more than seven orders of magnitude
between 400 and 200~nm (see Fig.~1).  The different regimes are commonly referred to as
UV-A (400--320 nm), UV-B (320--290 nm), and UV-C (290--200 nm).  The test
planets are assumed to be located in the stellar habitable zone (HZ).
Following the analyses by \cite{kas93} and \cite{und03}, we distinguish
between the conservative and generalized HZ (see Table~1).  For stars of
different spectral types, we also define planetary Earth-equivalent positions
given as $R_{\rm E} = \sqrt{L/L_\odot}$.  For the conservative HZ, the
inner limit of habitability is given by the onset of water loss that occurs
in an atmosphere warm enough to have a wet stratosphere, resulting in a
gradual loss of water by photodissociation and subsequent hydrogen loss to
space.  Moreover, the outer limit of habitability is given by
the first CO$_2$ condensation where for a surface temperature of 273~K, CO$_2$
begin to form.  Concerning the general HZ, the inner limit of habitability is
given by the runaway greenhouse effect.

In the following, we will present our results for different positions of
Earth-type planets while considering both stellar photospheric and different
levels of chromospheric radiation.  We will also discuss the likely
consequences of stellar flares.

\begin{figure*}
\begin{tabular}{cc}
\epsfig{file=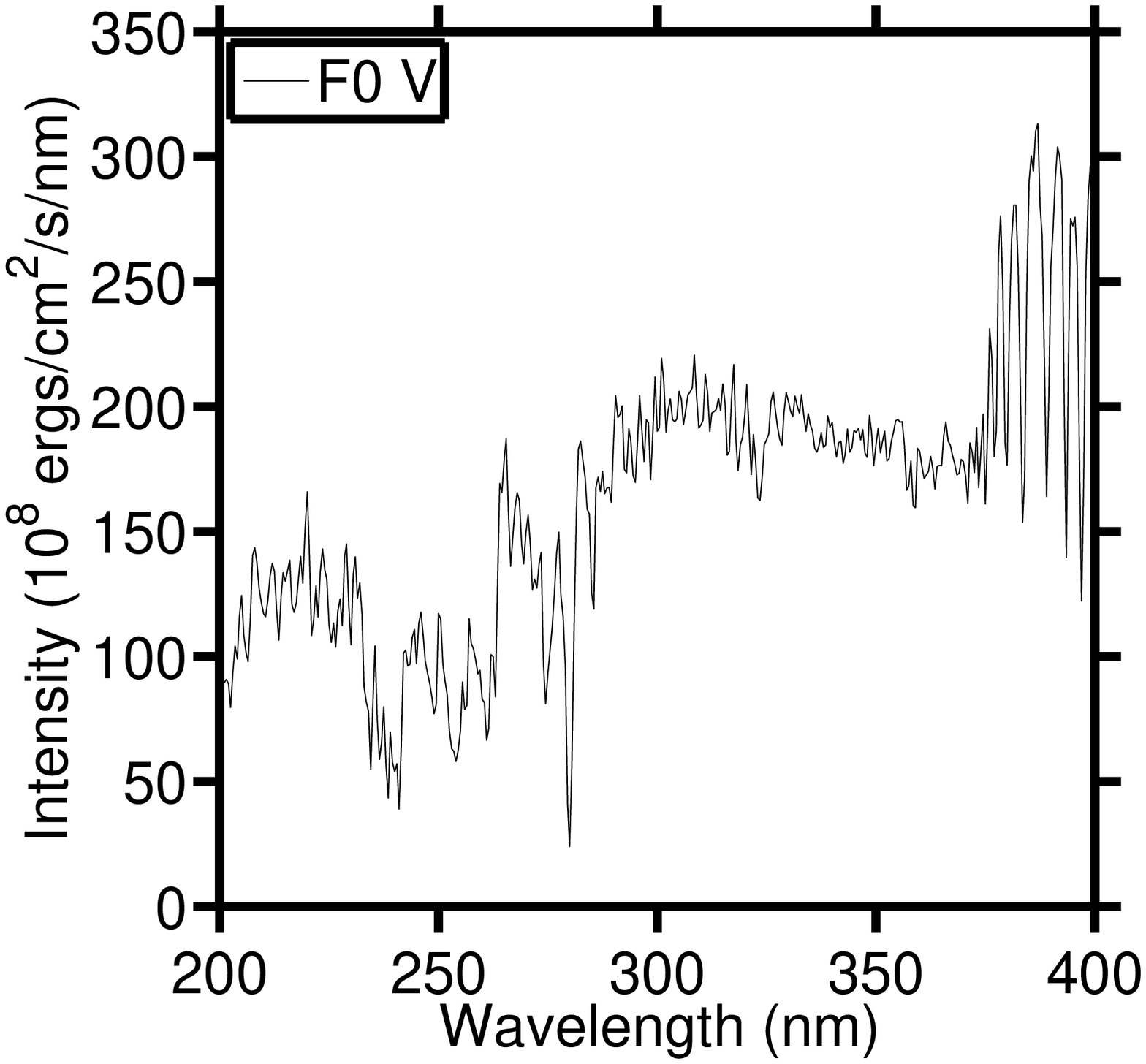,width=0.48\linewidth,height=0.38\linewidth} &
\epsfig{file=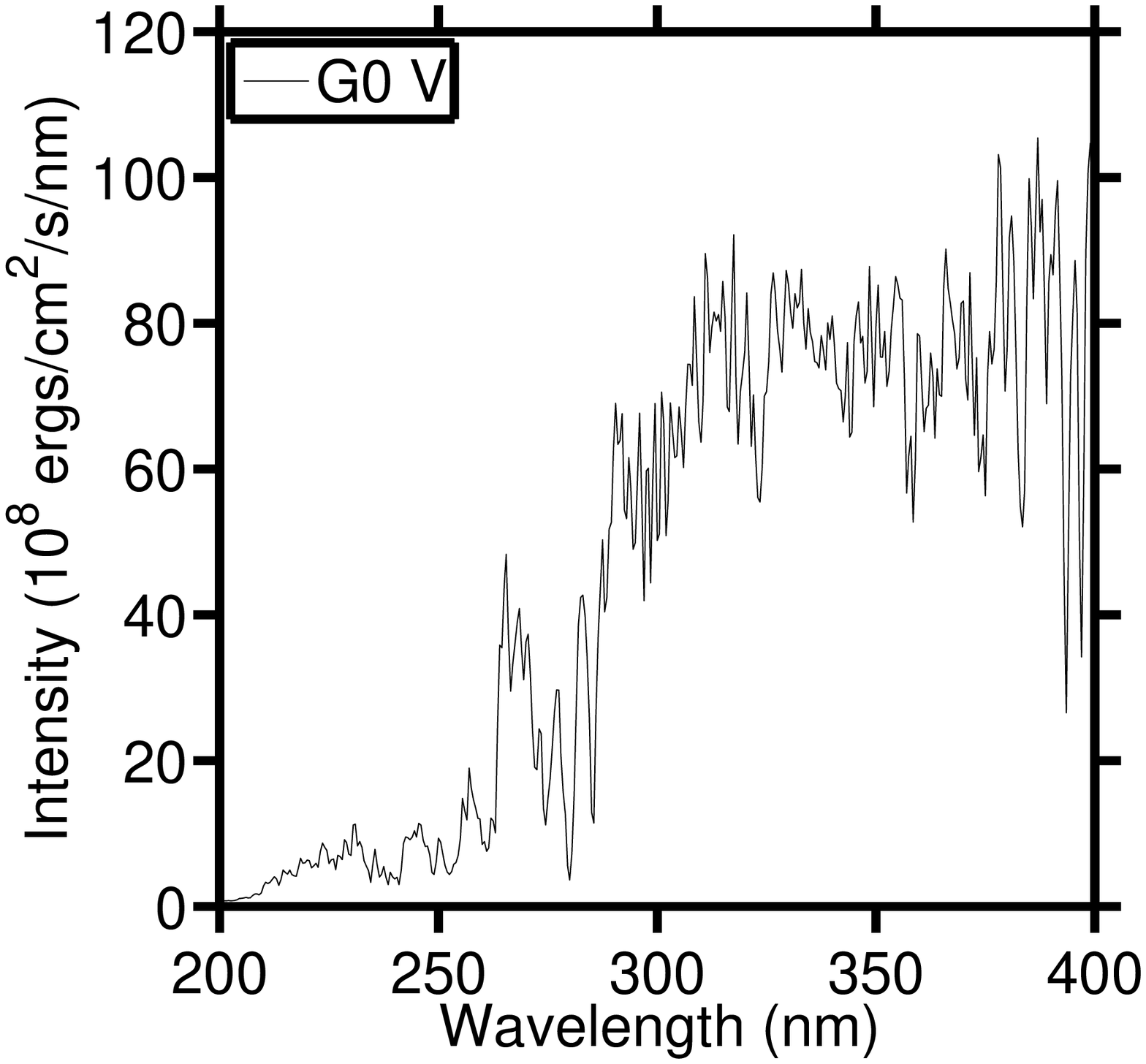,width=0.48\linewidth,height=0.38\linewidth} \\
\epsfig{file=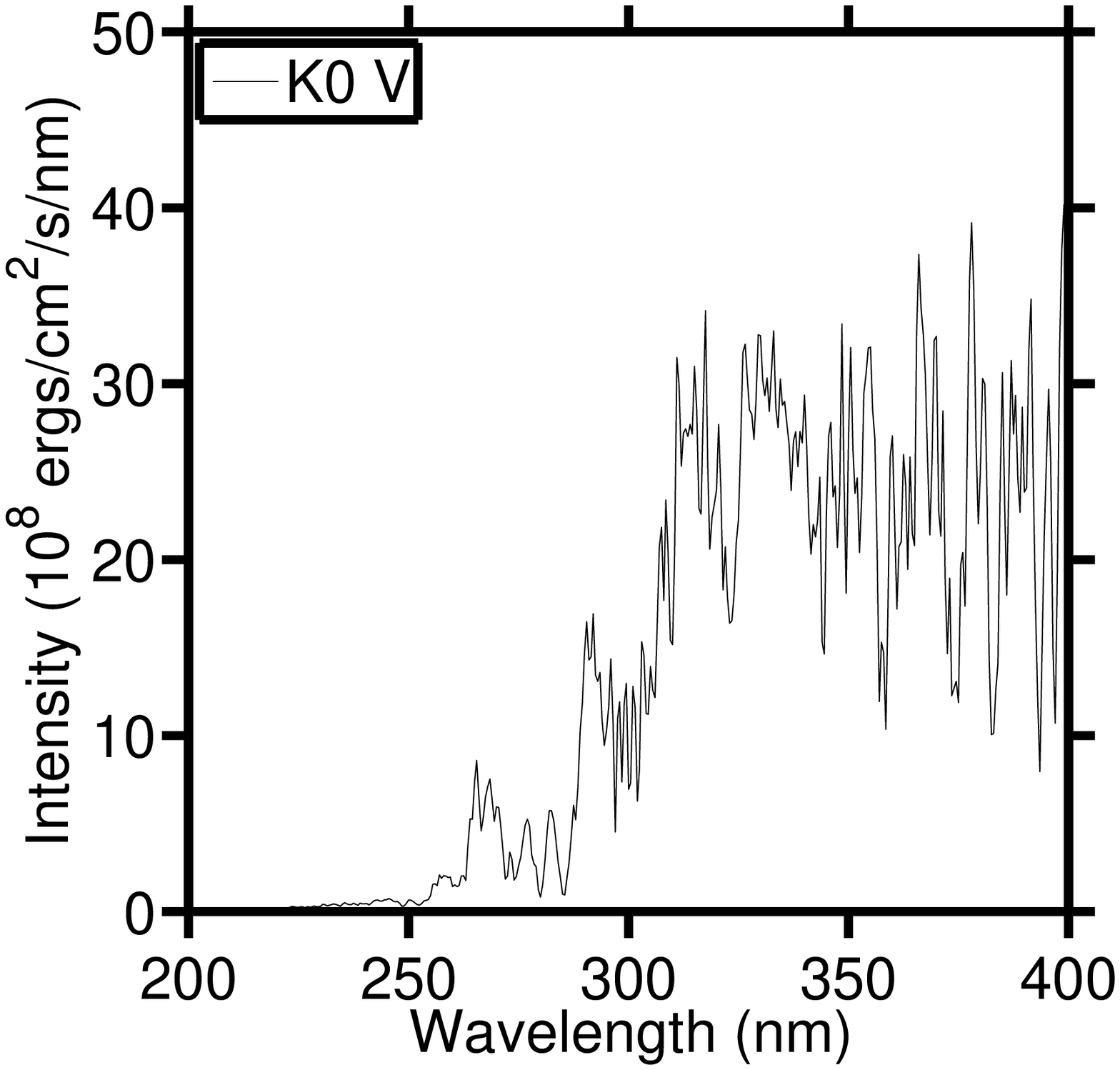,width=0.48\linewidth,height=0.38\linewidth} &
\epsfig{file=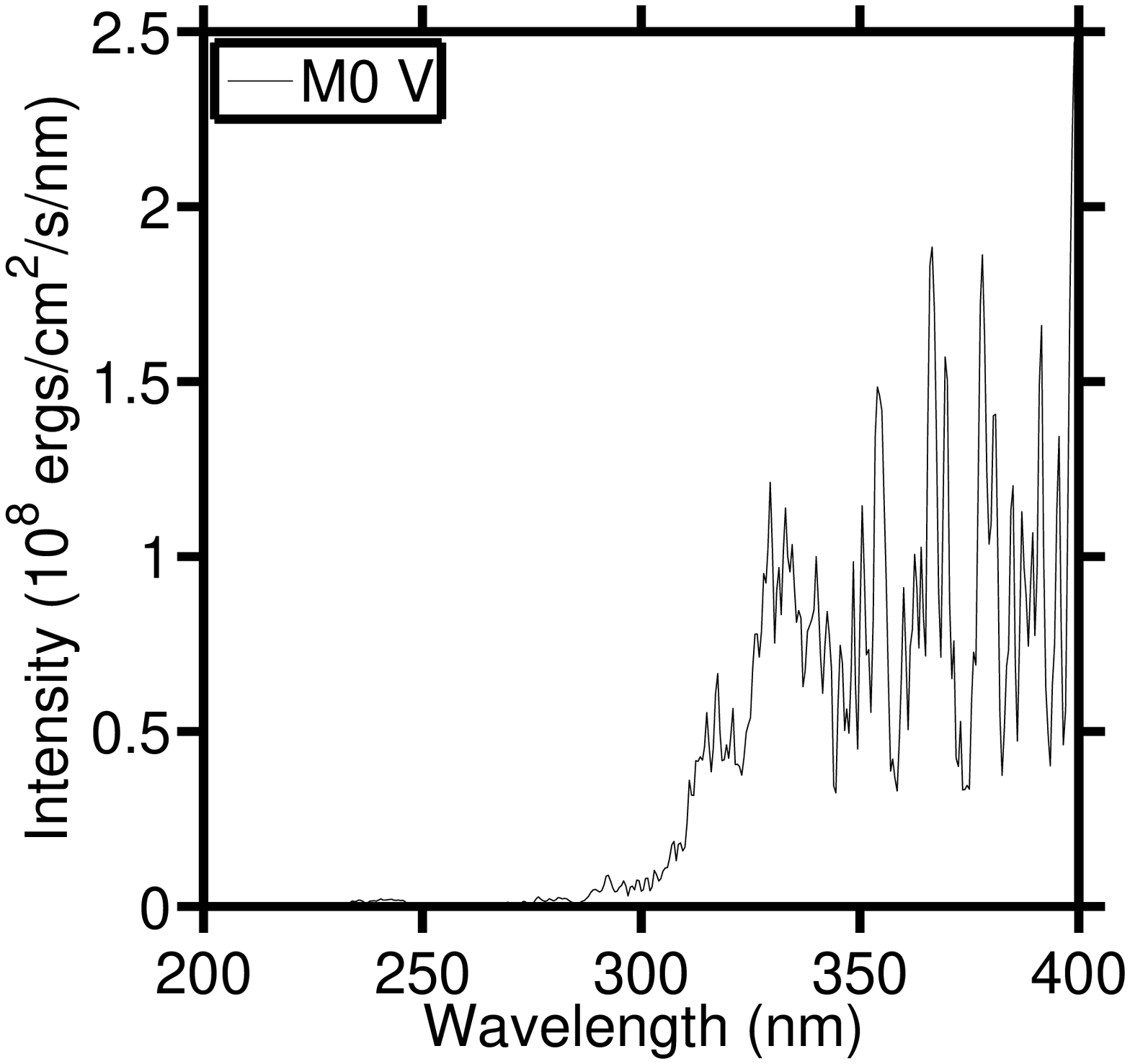,width=0.48\linewidth,height=0.38\linewidth}
\end{tabular}
\caption{Kurucz models for the F0~V, G0~V, K0~V, and M0~V stars used
in this study.  For the sake of display, the spectral resolution
has artificially been decreased to
${\lambda}/{\Delta\lambda} \simeq 300$, implemented by running means.
Note the different scales of the $y$-axes.
}
\end{figure*}

\begin{figure*}
\centering
\epsfig{file=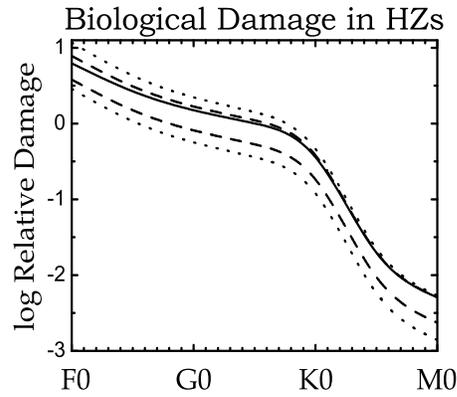,width=0.5\linewidth,height=0.45\linewidth}
\caption{Biological damage to DNA for a planet (no atmosphere) at an
Earth-equivalent position (solid line), and at the inner and outer limits
of the conservative (dashed lines) and generalized HZ (dotted lines).
}
\end{figure*}

\begin{figure*}
\centering
\epsfig{file=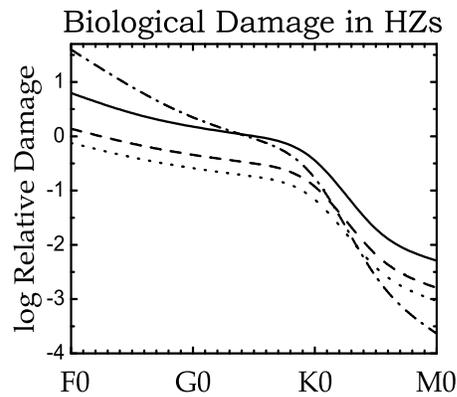,width=0.5\linewidth,height=0.45\linewidth}
\caption{Biological damage to DNA for a planet at an Earth-equivalent
position without an atmosphere (solid line), and an atmosphere akin to
Earth 3.5 Gyr ago (dashed line) and today (dotted line).  The dash-dotted
line refers to a planet without an atmosphere at 1~AU from its host star,
irrespectively of the position of the stellar HZ.
}
\end{figure*}

\section{Stellar Photospheric Radiation}

For our study we employ a detailed consideration of stellar
photospheric radiation for the UV-A, UV-B, and UV-C spectral regimes.  The
adopted target stars are: F0~V, F5~V, G0~V, G5~V, K0~V, K5~V, and M0~V
with effective temperatures of 7200, 6440, 6030, 5770, 5250, 4350, and 3850~K,
respectively (see Table~1).  As stellar surface gravity, log~g = 4.5 (cgs) is
used, and the microturbulence broadening is considered.  We make use of the
spectral models by R.~L.~Kurucz which take into account millions or hundred
of millions of lines for atoms and molecules; see \cite{cas04} and related
publications for details.

Note that it is virtually impossible to display the complete
photospheric spectra of our target stars owing to
the richness in their spectra features, particularly the very large
number of narrow lines due to the large number of elements and
line levels taken into account.  Thus, to provide a tutorial
comparison between the different spectra in the 200 to 400 nm wavelengths
regime, we created ``fake spectra" obtained by installing a running mean.
This has been done by smoothing the spectra for a length of 1~nm, corresponding
to artificially decreasing the spectral resolution to
${\lambda}/{\Delta\lambda} \simeq 300$ (see Fig.~2).

As part of our models, we also consider the effects of attenuation by
an Earth-type planetary atmosphere, which allows us to estimate
wavelength-dependent attenuation coefficients appropriate to the cases
of Earth as today, Earth 3.5 Gyr ago, and no atmosphere at all
(\cite[Cockell 2002]{coc02}).
Improved simulations of the effects of planetary atmospheres must be based
on the detailed treatment of atmospheric photochemistry, including the
build-up and destruction of possible ozone layers.

Our results are presented in Figs. 3 and 4.  They show the relative
damage to DNA due to different types of main-sequence stars encompassing
spectral spectral types between F0 and M0 normalized to today's Earth.
We consider Earth-type planets at the inner and outer edge of either
the conservative or generalized HZ as well as planets of
different atmospheric attenuation.  We obtained the following findings:
(1) All main-sequence stars of spectral type F to M have the potential of
damaging DNA due to UV radiation.  The amount of damage strongly depends
on the stellar spectral type, the type of the planetary atmosphere and
the position of the planet in the HZ; see \cite{coc99} for previous work.
(2) The damage to DNA for a planet in the HZ around an F-star (Earth-equivalent
distance) due to photospheric radiation is significantly higher (factor 5) compared
to planet Earth around the Sun, which in turn is significantly higher than for an
Earth-equivalent planet around an M-star (factor 180).
(3) We also found that the damage is most severe in
the case of no atmosphere at all, somewhat less severe for an atmosphere
corresponding to Earth 3.5 Gyr ago, and least severe for an atmosphere like
Earth today, as expected.  Moreover, any damage due to photospheric stellar
radiation is mostly due to UV-C.  The relative importance due to UV-B is
relatively small, and damage due to UV-A is virtually nonexistent
(see \cite[Cuntz et al. 2010]{cun10} for details).

\begin{table}
\caption{Increase in Biological Damage due to Chromospheric Radiation}
\centering
\begin{tabular}{c c c c}
\noalign{\smallskip}
\hline
\noalign{\smallskip}
Sp. Type & log~$F_{\rm basal}$ & Inact. Chr. & Act. Chr. \\
\hline
\noalign{\smallskip}
F0~V  &  5.890  &  ...  &  ...  \\
F5~V  &  5.755  &  ...  &  1.08 \\
G0~V  &  5.590  &  ...  &  1.12 \\
G5~V  &  5.486  &  ...  &  1.16 \\
K0~V  &  5.316  &  ...  &  1.4  \\
K5~V  &  4.879  &  1.6  &  6.7  \\
M0~V  &  4.498  &  2.6  &  17   \\
\hline
\noalign{\smallskip}
\multicolumn{4}{l}{Note: $F_{\rm basal}$ is given in ergs cm$^{-2}$ s$^{-1}$.} \\
\end{tabular}
\end{table}

\section{Effects by Stellar Chromospheres and Flares}

The study of chromospheric radiation and its effects on circumstellar
environments, including hosted planets, has been the focus of a large
array of research projects.  In our study, we added chromospheric
emission as exemplified by Mg~II {\it h+k} at 2803, 2796 {\AA} to
the photospheric emergent radiation.  Previous studies including those
associated with the ``Sun in Time" project (see Sect. 4), have shown
that for F and G-type stars, the photospheric UV continua are
typically dominant relative to any chromospheric contributions implying
that chromospherically induced biological damage for planets hosted
by these stars will be insignificant.

The situation is, however, decisively different for K and M dwarfs.
Any of these stars, including inactive stars, show noticeable
chromospheric emission.  Previous research (e.g., Cuntz et al. 1999,
and references therein) point to the existence of two-component
stellar chromospheres with the non-magnetic component (prevalent in
case of old stars; see Sect. 4) heated by acoustic waves and the
magnetic component heated by magnetic energy dissipation.  The
acoustically heated chromospheric component is usually also referred
to as basal component $F_{\rm basal}$, noting that the radiative
energy flux ranges from $3 \times 10^4$ (M dwarfs) to $8 \times
10^5$ ergs~cm$^{-2}$~s$^{-1}$ (F dwarfs); see \cite{rut91}.

Table 2 depicts the enhancement of biological damage due to
chromospheric radiation relative to photospheric radiation
for stars between spectral type F0~V and M0~V.
This study has been pursued for stars with both
basal (inactive) and significantly increased (active) chromospheric
emission.  For the latter, we considered an increase in chromospheric
emission by a factor of 10, which for solar-type stars amounts to
the maximal chromospheric emission observed in regard to
fast-rotating (i.e., young) stars (\cite[Vilhu \& Walter 1987]{vil87}).
Our results show that for
stars with basal chromospheric emission, chromosphere-induced
biological damage only occurs for stars of spectral type mid-K and
later, whereas for stars with relatively high chromospheric emission,
chromosphere-induced biological damage can also be found in the
environments of very late G-type and early K-type stars.

Other important agents for the delivery of highly energetic radiation,
especially UV-C, are flares (e.g., \cite[Pettersen 1989]{pet89},
\cite[Redfield et al. 2002]{red02}, \cite[Hawley et al. 2003]{haw03},
\cite[Robinson et al. 2005]{rob05}).
Note that flare activity is most pronounced in K and M-type stars.
It has also the potential of (partially) stripping the planetary
atmospheres (e.g., \cite[Kulikov et al. 2007]{kul07},
\cite[Lammer et al. 2008]{lam08}), which is especially relevant
for close-in planets, particularly planets located in the HZs of K and
M-type stars (see Table 1).  The maximal near-UV and far-UV radiative
energies of flares are akin or larger than those of an active stellar
chromosphere, although flare energy is typically provided episodically
rather than continuously.  Therefore, the results given in Table 2 may
also be applicable to flaring as a first approximation, except that
the actual biological damage due to flares might be enhanced further
by a factor of 5 to 10 due to the possible damage (i.e., evaporation)
of the planetary atmosphere.

Nonetheless a more detailed analysis of flare-related effects is
still required, which should also incorporate a detailed treatment
of planetary atmospheric photochemistry, including the build-up and
destruction of ozone as pointed out by
\cite[Segura et al. (2003, 2005)]{seg03,seg05} and others.

\section{Summary: The `Sun in Time' Project and Its Biology Expansion}

We attempted to quantify the biological damage expected
to occur in the environments of late-type main-sequence stars due to
photospheric, chromospheric and flare radiation.  Stellar photospheric
radiation has been considered by utilizing spectral models given by
R.~L.~Kurucz and collaborators, which take into account millions or
hundred of millions of lines for atoms and molecules; see, e.g.,
\cite{cas04}.  Concerning chromospheric radiation, we considered models
exemplifying basal chromospheric emission (e.g., \cite[Schrijver 1987]{sch87},
\cite[Rutten et al. 1991]{rut91}) and significantly increased
chromospheric emission (e.g., \cite[Vilhu \& Walter 1987]{vil87}).
Note that basal emission is usually attributed to acoustic heating
in the limiting case of magnetic energy dissipation to be small or
negligible.  Detailed models of one-component and two-component
chromospheric heating for late-type stars have been given by,
e.g., \cite{buc98}, \cite[Cuntz et al. (1998, 1999)]{cun98,cun99},
\cite{faw02}, and \cite{ram03}.

It is highly noteworthy that there is a deeper underlying connection
between the level of chromospheric heating and emission, on one
hand, and more fundamental stellar physics involving processes
concerning the stellar interior and phenomena associated with
stellar winds, on the other hand.  In case of the Sun, this
connection has been explored in detail by, e.g., \cite{gui02},
\cite{gui03}, \cite{rib05}, and \cite{gue07}.  A highly crucial
question, as gauged by its implied planetary, biological and
societal percussions, is whether the Sun has always been a
relatively inactive star or, in contrast, has experienced some
periods of stronger magnetic activity.

Compelling observational evidence (\cite[G\"udel et al. 1997]{gue97})
shows that zero-age main-sequence (ZAMS) solar-type stars rotate over
10 times faster than today's Sun.  As a consequence of this, young
solar-type stars, including the young Sun, have vigorous magnetic
dynamos and correspondingly strong high energy emissions.  From
the study of solar-type stars of different ages, \cite{sku72},
\cite{sim85}, \cite{cha93}, \cite{mac94} and others showed that
the Sun loses angular momentum with time via magnetized winds
(magnetic breaking), thus leading to a secular increase in its
rotational period (e.g., \cite[Durney 1972]{dur72},
\cite[Keppens et al. 1995]{kep95}).  This rotation slowdown is well
fitted by a Skumanich-type power law roughly proportional to $t^{-1/2}$
(e.g., \cite[Skumanich 1972]{sku72}, \cite[Soderblom 1982]{sod82},
\cite[Ayres 1997]{ayr97}).  In response to slower rotation,
the solar dynamo strength diminishes with time, causing the
Sun's high energy emissions and mass loss also to undergo
significant decreases.

A direct consequence of this behavior concerns the generation of
photospheric and chromospheric magnetic flux, which constitute the
physical reason for the varying level of chromospheric heating and
associated generation of UV and EUV radiation with stellar age or
rotation period (e.g., \cite[Mathioudakis et al. 1995]{mat95}).
Consequently, it is thus also possible to relate the photospheric
magnetic flux to the stellar rotation period
(\cite[Noyes et al. 1984]{noy84}; \cite[Marcy \& Basri 1989]{mar};
\cite[Montesinos \& Jordan 1993]{mon93}; \cite[Saar 1996a]{saa96a})
as well as to the emergent chromospheric emission 
(\cite[Saar \& Schrijver 1987]{saa87}; \cite[Schrijver et al. 1989]{sch89};
\cite[Montesinos \& Jordan 1993]{mon93}; \cite[Saar 1996b]{saa96b};
\cite[Jordan 1997]{jor97}).

The acquisition, interpretation and modeling of data for any of
these phenomena is the main focus of the ``Sun in Time" project
(originally solely focused on G0--G5~V stars) pursued under the
leadership of E.~F. Guinan, which relates, in a statistical
sense, the strength of the emergent UV, EUV and X-ray flux and the
stellar mass loss rate to the stellar age.  More recently, this
project has been extended to K-type and M-type dwarfs with the
latter element of study now referred to ``Living with a Red Dwarf".
The critical ultimate connection to be targeted in the future, see
\cite{gui09} for the dissemination of various intermediate results,
is to relate the type and strength of stellar activity for
different types of stars to planetary climatology, particularly for
planets in stellar HZs, and to exosolar planetary biology, if existing.

\begin{acknowledgments}

M.C. would like to thank the American Astronomical Society and the IAU for
travel support.

\end{acknowledgments}

\end{document}